# Effect of divalent Ba cation substitution with Sr on coupled 'multiglass' state in the magnetoelectric multiferroic compound $Ba_3NbFe_3Si_2O_{14}$

**Satyapal Singh Rathore and Satish Vitta***
Indian Institute of Technology Bombay
Mumbai 400076; India.

## Abstract

$(Ba/Sr)_3NbFe_3Si_2O_{14}$ is a magneto-electric multiferroic with an incommensurate antiferromagnetic spiral magnetic structure which induces electric polarization at 26 K. The structure, as revealed by x-rays and neutrons, as well as static and dynamic magnetic and dielectric properties of these compounds have been studied down to 6 K under different conditions. Both the compounds have similar crystal structure but with different lattice constants down to 6 K. The Ba-and Sr-compounds exhibit a transition at 26 K and 25 K respectively, as indicated by the specific heat capacity and dc specific magnetization, into an antiferromagnetic state. Although Ba and Sr are isovalent, they exhibit very different static and dynamic magnetic behavior. The Ba-compound exhibits both thermal and magnetic field hysteresis with the thermal hysteresis decreasing with increasing magnetic field, a behavior typical of glasses. The glassy behavior is also clearly seen in the ac susceptibility studies which show a dispersive peak in the range 40 K to 90 K in the frequency range $10^1$ Hz to $10^4$ Hz. The dispersive behavior follows a cluster glass critical slowing dynamics with a freezing temperature of ~ 35 K and a critical exponent of 3.9, a value close to the 3-D Ising model. The coercive field in the Ba-compound is found to increase from 1700 Oe to 2500 Oe with increasing temperature with an exchange bias field of 82 Oe at 25 K which reduces above the transition to reach a negative value of 178 Oe at 200 K before vanishing at room temperature. The Sr-compound however does not exhibit any dispersive behavior except for the invariant transition at 25 K in ac susceptibility with no magnetic field hysteresis at all temperatures. The dielectric constant studied in the frequency range $10^1$ Hz to $10^6$ Hz also reflects the magnetic behavior of the two compounds. The Ba-compound has two distinct dispersive peaks – near $T_N$ and in the range 40 K to 125 K while the Sr-compound has a single dispersive peak in the range 40 K to 80 K. The activation energy of the high temperature dispersive peak in both compounds however is found to be similar, 71 meV and 65 meV respectively for Ba- and Sr-compounds. These results clearly show that the magneto-electric behavior in these compounds is extremely sensitive to the nature of super exchange paths and can lead to electronic phase separation exhibiting complex behaviors.

**Introduction:**

The search for materials wherein the magnetization M and electric polarization P are strongly coupled has intensified recently with the discovery of magneto-electric coupling in a variety of materials.[1-5] In these materials both space inversion and time reversal symmetry need to be broken to realize coupling between the two orders. The linear magnetoelectric coupling constant $\alpha_{ij}$, given by $\sqrt{\varepsilon_{ii}^2 \mu_{jj}^2}$ depends on the magnetic permeability tensor $\mu_{jj}$ and electric permittivity tensor $\varepsilon_{ii}$ and requires that they be as large as possible for 'strong coupling'.[6] However, most materials have either small $\mu's$ or small $\varepsilon's$ or both and hence the coupling constant $\alpha_{ij}$ becomes even smaller. Hence in the search for materials with strong coupling, materials with non-conventional magnetic structures such as spirals and toroids have been investigated and they have been found to exhibit strong coupling.[7-11] An alternative approach to overcome the limitations of linear coupling would be to investigate materials wherein higher order quadratic coupling constants become dominant compared to the linear coupling constant. This can be seen from the total free energy F expression of a homogenous, stress-free material given by;

$$F(\boldsymbol{E}, \boldsymbol{H}) = -\left[\left(\frac{\varepsilon_0 \varepsilon_{ij} E_i E_j}{2}\right) + \left(\frac{\mu_0 \mu_{ij} H_i H_j}{2}\right) + \left(\alpha_{ij} E_i H_j\right) + \left(\frac{\beta_{ijk} E_i H_j H_k}{2}\right) + \left(\frac{\gamma_{ijk} H_i E_j E_k}{2}\right) + \cdots\right] \text{ Eq. (1)}$$

where $\varepsilon_0$ $and$ $\mu_0$ refer to free space permittivity and permeability respectively, $H$ and $E$ are magnetic and electric fields. An advantage of invoking the quadratic terms is that the system need not fulfill the symmetry criteria. Disordered multiferroics which do not fulfill the stringent symmetry criteria of linear magnetoelectric coupling belong to this class of materials and have received considerable attention in recent years. Materials with complex magnetic structure with multiple degenerate ground states are another class of materials which can have large higher order quadratic coupling constants.[12-16]

Recently, $Ba_3NbFe_3Si_2O_{14}$, a langasite family compound, has been shown to exhibit a frustrated, spirally ordered antiferromagnetic (AF) structure which leads to ferroelectric (FE) ordering at 26 K. The magnetic $Fe^{3+}$ (S=5/2) ions have a triangular coplanar arrangement in the a-b plane and form a trimer with a frustrated spin structure, Figure 1. The equilateral triangular structure of $Fe^{3+}$ ions has been found to get deformed for T < $T_N$ leading to the formation of a lower symmetry polar C2 or P3 structures compared to the non-polar P321 structure present above $T_N$. This trimer structure has a helical arrangement along c-axis and orders antiferromagnetically below 26 K, the Neel temperature, $T_N$. The magnetic spiral is known to be commensurate with helical period of 3.66 nm, 7 times the lattice parameter along c-direction leading to an angular rotation of $51^0$ of the magnetic moments between the different a-b planes. The helical magnetic structure is stabilized by the two in-plane ($J_1$, $J_2$) and three out-of-plane ($J_3$, $J_4$, $J_5$) Fe-O-O-Fe super-exchange interactions. This magnetic structure has been found to trigger formation of a polar lattice for $T < T_N$ and hence ferroelectricity.[17,18] The macroscopic properties of this compound however have been found to be far more complex compared to the simple structural perspective mentioned above. The magnetic contribution to the specific heat $C_P$ was found to rise significantly above the background value at temperatures as high as 100 K, 4 $T_N$, indicating that ~ 40 % of the magnetic transition entropy is released at $T \gg T_N$ and that the magnetic ordering begins far above $T_N$.[19] The magnetic structure investigated using neutrons clearly showed the presence of short range magnetic order up to ~ 100 K and that a truly long range AF structure developed only for $T \leq 10K$.[20] The recent ESR studies indicate that the low T long range left handed magnetic spiral structure is established by the anisotropic Dzyaloshinskii-Moriya (DM) interactions between the $Fe^{3+}$ moments with interaction energies of ~ 120 mK and 45 mK perpendicular and parallel to the c-axis respectively.[21,22] Absorption studies performed using linearly polarized THz synchrotron radiation further indicate the presence of magnons at low T and phonons dressed with currents at $T \gg T_N$. These phonons exhibit a magnetic behaviour which is not due to atomic spins and a helicoidal polarization state which is incommensurate with the crystal lattice is proposed to exist even at $T > T_N$.[23] This has been confirmed by recent polarization studies which show the presence of intrinsic electric polarization in the compound. The intrinsic polarization present in the absence of a magnetic field has been found to reverse its direction with increasing magnetic field.[24] Co-existence of multiple magnetic states, ferromagnetic, antiferromagnetic and glassy under different conditions has also been observed recently.[25]

The above discussed studies clearly show that $Ba_3NbFe_3Si_2O_{14}$ with a frustrated helical antiferromagnetic ground state exhibits non-trivial magnetic orders at different temperatures and also shows a tendency for electronic phase separation. There is however a clear lack of investigations of macroscopic behaviour in detail. Hence in the present work, the magnetic and electric polarizations have been studied in detail using both static and dynamic techniques. These are corroborated with investigation of the magnetic structure down to 6 K using neutrons. Since the polycrystalline form of $Ba_3NbFe_3Si_2O_{14}$ has been investigated, the effect of various structural defects including that of antisite defects is revealed in these studies. The spiral magnetic structure is a result of both in-plane and out-of-plane $Fe^{3+}$ ionic interactions which depends strongly on the spatial coordinates of various cations and $O^{2-}$. The divalent cation, $Ba^{2+}$ which plays a crucial role in out-of-plane magnetic interactions has been substituted with a smaller $Sr^{2+}$ cation to study its effect on the magnetic structure and hence macroscopic properties. It is found that substitution changes the super exchange path length and more importantly the magnetic and dielectric properties both below and above the transition temperature.

**Experimental details:**

For synthesizing $(Ba/Sr)_3NbFe_3Si_2O_{14}$ (BNFSO/SNFSO) compounds, solid state reaction technique was employed as described in Ref. [25], with high purity $BaCO_3$, $Nb_2O_5$, $Fe_2O_3$, $SiO_2$ and $SrCO_3$ as starting materials. The powders were mixed in a ball mill and then calcined several times at 1100 $^0$C to get a homogenous single phase. This single-phase powder was then pressed into a pellet and sintered at 1150 $^0$C to get a pellet with > 95% density. These sintered pellets were used for all structural and properties characterizations. The structure and chemical composition were determined using a combination of X-ray and neutron diffraction, electron microscopy and x-ray photoelectron spectroscopy (XPS). X-ray diffraction was performed using Cu-K$_\alpha$ radiation of wavelength 0.154 nm, while neutron diffraction was performed with 0.12443 nm wavelength neutrons as a function of temperature down to 6 K at the Dhruva reactor, Bhabha Atomic Research Centre. XPS was done using Mg Kα X-rays and the spectra were analysed using the standard database.[26] The specific heat capacity, $C_p$ was measured both as a function of temperature T and magnetic field H up to 5 T in a steady state relaxation mode using physical property measurement system. The magnetization M and susceptibility χ were measured in the temperature range 5 K to 300 K, fields up to ± 3 T and frequency range $10^1$ to $10^4$ Hz using SQUID based magnetic property measurement system and vibrating sample magnetometer. The ac susceptibility was measured using 2.5 Oe excitation field without any dc field bias. The variation of dielectric constant, $\epsilon^*$ as a function of varying temperature, 5 K to 300 K and frequency, $10^1$ to $10^6$ Hz in the presence of 0.5 V potential was studied using Novocontrol frequency–response analyser.

**Results:**

**X-ray and Neutron Diffraction:**

The room temperature X-ray diffraction patterns obtained from $Ba_3NbFe_3Si_2O_{14}$ (BNFSO) and $Sr_3NbFe_3Si_2O_{14}$ (SNFSO) are shown in Figure 2(a) along with the corresponding neutron diffraction pattern. All the reflections in both the compounds could be identified and indexed to the non-centrosymmetric non-polar hexagonal structure P321 indicating absence of any other phase. These results are in complete agreement with earlier studies performed with single crystals.[17] Since the ionic size of $Sr^{2+}$ is less than that of $Ba^{2+}$, this could result in variation of lattice parameters. In order to determine the changes in structural parameters, Rietveld refinement of the structure was performed and the data is fitted to the experimental pattern. The results are shown in Figure 2(a) and the crystallographic data obtained after refinement are given in Table 1. It is seen that both the lattice parameters, 'a' and 'c' indeed change and were found to decrease due to substitution of $Ba^{2+}$ with $Sr^{2+}$. The in-plane lattice parameter 'a' decreases from 0.85001 nm to 0.82605 nm while the out-of-plane lattice parameter 'c' decreases from 0.52267 nm to 0.51298 nm. The microstructure as seen in a scanning electron microscope, Figure 2(b) shows the presence of large grains with average grain size of 5 μm and no secondary phases. The chemical composition as well as the ionic state of all the different ions present in the compound was determined by XPS and the results are shown in Figure 3. All the peaks in the survey spectrum could be identified to the five elements Ba, Nb, Fe, Nb, Si and O indicating the purity of the compound. The core level spectra show that all the ions are in the expected ionic state, i.e. $Ba^{2+}$, $Sr^{2+}$, $Nb^{5+}$, $Fe^{3+}$, $Si^{4+}$ and $O^{2-}$. The presence of Fe in only the 3+ state clearly confirms the high spin, S=5/2 state of the compounds which is also seen in the room temperature X-band electron spin resonance spectrum, Figure 3 (g). A single resonance peak is seen at a magnetic field of 3394 Oe. The gyromagnetic ratio 'g' determined using the resonance field is found to be 2.0 corresponding to the spin only magnetic behaviour of the compound. In order to check for structural stability/changes that can occur with changing T, temperature dependent elastic neutrons diffraction studies of $Ba_3NbFe_3Si_2O_{14}$ were undertaken and the results are shown in Figure 4(a). All the peaks in the diffraction pattern at all temperatures across the phase transition could be identified to the trigonal structure and the lattice parameters determined are shown in Figure 4(b). The two lattice parameters 'a' and 'c' vary continuously with 'a' decreasing while 'c' increasing with temperature T. The two lattice parameters however show a discontinuous change at $T_N$ with a peak indicating a change in structural symmetry to a lower ordered structure.[27] Additional peaks corresponding to the magnetic order could be detected only below $T_N$ indicating that the long range magnetic structure develops only for $T < T_N$ with short range magnetic structural entities being present above $T_N$. The presence of short range magnetic correlations in BNFSO up to T as high as 100 K have only been observed in diffuse scattering of neutrons studies[20] and hence could not be detected in the present studies. The magnetic structure obtained by refinement of the 6 K diffraction data shows an incommensurate ordering of $Fe^{3+}$ moments along the c-direction with the propagation vector κ(0,0,0.1441) corresponding to helical ordering of the magnetic moments with a period of ~ 7c. The effective magnetic moment is found to be ~ $4.2\mu_B$ per $Fe^{3+}$ ion, in agreement with earlier studies.[17,27] The lower value of magnetic moment is due to the partial covalent nature of $Fe^{3+}$ electrons in the tetrahedral configuration of $O^{2-}$ ions.

**Heat Capacity:**

The heat capacity $C_P$ measured both as a function of temperature T in the range 5 K to 300 K and magnetic field H up to 5T is shown in Figure 5. The heat capacity shows a clear peak at all fields at 26 K for BNFSO and at 25 K for SNFSO corresponding to antiferromagnetic phase transition $T_N$ in both the compounds. The heat capacity is found to be nearly independent of magnetic fields at all temperatures i.e. the zero field and field dependent $C_p$ overlap at all temperatures. An additional feature noticed is the presence of a broad hump in the temperature range 40 K to 100 K in both the compounds and is attributed to onset of ordering with the consequent release of entropy.[19] The magnetic contribution to the heat capacity however cannot be extracted without knowing the exact non-magnetic or phonon contribution. Since the compositions studied in the present work all have $Fe^{3+}$ this could not be done. The absence of magnetic field dependence of $C_p$ in both BNFSO and SNFSO indicates that the spin configuration does not change with changing field and that the magnetic order is inherent to the system. The strong temperature dependence as well as the hump like feature in the heat capacity indicate that both correlated as well as orphan spins influence the heat capacity and the orphan spins could be due to the presence of a variety of structural defects which can result in the formation of magnetically different phases.[28,29]

**Magnetization (dc and ac):**

The dc magnetization also exhibits a clear field independent cusp at 26 K and 25 K respectively for BNFSO and SNFSO corresponding to the antiferromagnetic transition as seen in Figure 6 and is in agreement with heat capacity data. The similarity between the two compounds BNFSO and SNFSO; and the earlier reported studies[30] on BNFSO however ends here. The magnetic field and path dependence of the magnetization M, i.e. zero field cooling and field cooling, exhibits clear irreversibility in the case of BNFSO which is absent in SNFSO. The irreversibility temperature $T_{irr}$ decreases with increasing magnetic field and becomes '0' for fields > 5000 Oe. These results indicate the presence of a 'spin-glass' like phase in BNFSO which is not observed in SNFSO. The magnetic field dependence of $T_{irr}$ in the case of BNFSO however does not follow the de-Almeida-Thouless phase boundary relation,[31] indicating an unconventional 'glassy' behaviour. The high temperature magnetization in the case of BNFSO follows Curie-Weiss behaviour only for T > 200 K and an extrapolation gives a value of ~ – 350 K for $\theta_{CW}$ the Curie-Weiss temperature compared to ~ – 90 K for SNFSO. These values show that the level of geometric frustration in the case of BNFSO is much greater than that present in SNFSO, ~ 3.6. The atomic magnetic moment determined from the Curie constant for BNFSO is found to be ~ 5.73 $\mu_B$ compared to 5.79 $\mu_B$ for SNFSO. The lower than expected magnetic moment, 5.92 $\mu_B$, is attributed to the frustration present in the structure and the extent of covalency in the $Fe^{3+}$ electrons. Hence, it can be concluded that the additional feature observed in the heat capacity and the irreversibility in magnetization in the case of BNFSO are solely a consequence of short range magnetic effects such as formation of 'spin-glass' like phase. The presence of 'spin-glass' phase in the case of BNFSO is further confirmed by ac magnetic susceptibility studies discussed in the next paragraph. In the case of SNFSO however the short range ordered magnetic entities do not seem to affect the dc magnetization.

The frequency ω and temperature T dependence of ac magnetic susceptibility, shown in Figure 7(a) shows two clear peaks – a ω-independent peak at 26 K and a broad dispersive peak in the temperature range 40 K to 80 K. The ω-independent peak at 26 K corresponds to the magnetic phase transition into an antiferromagnetic state and is 'athermal' in nature. The second peak in the range 40 K to 80 K however is highly dispersive and shifts to higher temperatures with increasing ω. The width of the peak decreases with decreasing frequency, a characteristic feature of spin glass systems. The shift of peak temperature $T_p^m$ with frequency ω given by the factor $\varphi_m = \Delta T_p^m / T_p^m \Delta \log(\omega)$ is generally used to qualitatively determine the type of magnetic behaviour.[32] In the present case of BNFSO it is found to be ~ 0.2, a value higher than that corresponding to a prototypical spin glass but much smaller than that for a superparamagnetic system. The magnetic freezing temperature $T_f^m$ for crossover from a paramagnetic state to a glass with critical slowing spin dynamics can be obtained using the relation;

$$\omega_{max}^m \propto |\epsilon|^{z\gamma} \qquad \text{Eq. (2)}$$

where $\epsilon$ represents the reduced temperature $\left(T_p^m/T_f^m - 1\right)$ and $z\gamma$ is the critical exponent. The magnetic freezing temperature $T_f^m$ is found to be 34.7 K, indicating that the high temperature glassy clusters phase is frozen above $T_N$ and this phase further transforms to the antiferromagnetic phase at 26 K. The critical exponent is found to be 3.9, corresponding to the 3D Ising model[32] with a slowing time for collective particles relaxation of 1.2x10$^{-4}$ s. These results clearly show that the dispersive magnetic susceptibility is due to the formation of a cluster glass phase. The ac magnetic susceptibility in the case of SNFSO, Figure 7(b), however is very different compared to that observed in BNFSO. The susceptibility exhibits a single frequency independent peak at 25 K, $T_N$, with no other high temperature peak. This clearly shows that SNFSO undergoes just one phase transition from the paramagnetic state to antiferromagnetic state, in contrast to the behaviour observed in BNFSO where in this transition is mediated via the glassy state.

The isothermal variation of magnetization M with magnetic field H at temperatures both below and above $T_N$ has also been investigated and the results are shown in Figure 8 for both BNFSO and SNFSO. The magnetization does not saturate in both the cases for fields as large as 3 T, clearly showing the paramagnetic/antiferromagnetic nature above and below $T_N$ respectively. The magnetization in the case of BNFSO is clearly due to at least two distinct magnetic phases at any temperature, below and above $T_N$. This can be seen clearly in the derivative plot shown in Figure 8(b). The derivative of magnetization $\partial M/\partial H$ is highly non-linear with peaks at low fields and tends to have a constant value at large fields – behavior typical to a mixture of ferromagnetic and paramagnetic/antiferromagnetic phases. The magnetization exhibits a clear coercive field $H_c$ which increases from 1700 Oe to 2500 Oe with increasing temperature in the entire temperature ranging from 5 K to 300 K. It is to be noted that the hysteresis behavior also exhibits considerable exchange bias with a peak in the bias field at $T_N$ which reverses sign both below and above $T_N$ clearly showing the presence of antiferromagnetic phase together with a ferromagnetic phase, Figure 8(a) inset. The low temperature biasing field sign reversal takes place between 10 K and 20 K while that above $T_N$ is around 50 K and vanishes at room temperature. The reversal of sign of the biasing field with temperature clearly indicates the complex nature of the magnetic structure as well as the magnetic transitions in the case of BNFSO. The low T crossover is plausibly due to the formation of a long range ordered antiferromagnetic phase while that at high temperatures is due to formation of spin glass clusters. The dc magnetization and ac susceptibility show that multiple phases, antiferromagnetic, ferromagnetic, spin glass and paramagnetic, coexist in BNFSO in different temperature ranges. The substitution of $Ba^{2+}$ with an isovalent $Sr^{2+}$ changes the magnetic response completely and does not exhibit any anomalous behavior as seen in Figure 8(c). The magnetization does not exhibit any hysteresis indicative of the presence of a single phase, paramagnetic above $T_N$ and antiferromagnetic below $T_N$. It should however be noted that below $T_N$ the system has a non-linear magnetic field response which changes to a near linear response above $T_N$ typical of an ideal paramagnet, Figure 8(d). The high temperature magnetization is linear at all temperatures with the susceptibility obeying the Curie-Weiss behavior down to 100 K with a magnetic moment of ~ 5.8 $\mu_B$ in complete agreement with the magnetic moment determined in the case of BNFSO.

**Dielectric Constant:**

The dielectric constant ε studied both as a function of frequency ω in the range $10^1$ Hz to $10^5$ Hz and temperature T for both BNFSO and SNFSO is shown in Figure 9. The dielectric constant shows distinctly different behavior between the two compounds. In the case of BNFSO two clear dispersive peaks are observed in the temperature range 20 K to 35 K (peak I) and 40 K to 125 K (peak II), Figure 9(a) while SNFSO exhibits a single high temperature dispersive peak in the range 40 K to 80 K, Figure 9(b) with a frequency independent non-dispersive hump at $T_N$. The dispersive dielectric response observed in both the cases is similar to that observed in relaxor ferroelectrics wherein polar nano regions (PNR) have been known to form due to displacement of the ions from their equilibrium positions. The PNRs are known to be extremely stable and robust with extremely large fields being required to destroy them.[33] The frequency dependence of the peak temperature $T_p^e$ defined by the parameter $\varphi_e$, similar to that used to define the magnetic susceptibility dispersion is found to be ~ 0.15 for peak I and ~ 0.2 for peak II in the case of BNFSO. These parameters are similar to that determined from ac susceptibility studies and clearly illustrate the magnetic and electric similarities and hence a direct coupling between the two order phenomenon. The temperature dependence of frequency maximum for both the peaks and in both the compounds follows a Vogel-Fulcher type relaxation given by;

$$\omega_{max}^e = \omega_0 exp[-E_A/k_B(T_p^e - T_f^e)] \qquad \text{Eq. (3)}$$

where $\omega_0$ is the attempt frequency, $E_A$ the barrier activation energy and $T_f^e$ the glass freezing temperature. The activation barrier at low temperature is found to be 40 meV while that for high temperature is ~ 71 meV in the case of BNFSO. The activation energy in the case of SNFSO is ~ 65 meV, similar to that determined for BNFSO peak II. The activation energy determined for peak I in the case of BNFSO is similar to that observed in the case of $Pr_{0.7}Ca_{0.3}MnO_3$ wherein the polarization is phonon assisted and results in the formation of polaronic charge carriers at low temperatures.[34] The low temperature freezing temperature for BNFSO is found to be 1.8 K, in agreement with low temperature magnetic structural studies which show the presence of short range clusters till ~ 1.5 K.[35] The freezing temperature $T_f^e$ for the high temperature peak in BNFSO however is found to be 25 K, close to $T_N$, while in the case of SNFSO this temperature reduces to 3.8 K indicating that the polar glass phase is present even below $T_N$. This difference in freezing temperatures is due to the intervening phase transition into a second relaxor state at 26 K in the case of BNFSO. The formation of polar nano regions in these relaxors is attributed to fluctuating dipoles with the nano regions having lower symmetry compared to the surrounding matrix in which they are embedded.[36] The scale of these domains however is very small and is predicted to increase with decreasing temperature, thus making it extremely difficult to observe them experimentally.

**Discussion:**

Materials with frustrated antiferromagnetic interactions coupled with commensurate/incommensurate helical spiral structure are unique and exhibit a variety of unusual magnetic structures. The existence of energetically favorable multiple ground states in such materials can also lead to co-existence of multiple phases ranging from ferromagnetic to antiferromagnetic and various types of glassy states. The co-existence of multiple phases is often driven by kinetic constraints which lead to arresting of phase transformations. This results in the formation of non-ergodic structures wherein ordered and disordered phase in-homogeneities can coexist and can be at the nanoscale or at longer mesoscopic length scales.[37] The magneto-electric coupling in such materials can exhibit unique features including giant higher order coupling constants which makes them extremely promising for multiferroic applications.[38] The ideal ground state in the case of BNFSO is a frustrated trimer with an incommensurate helical magnetic structure along the c-direction. This structure has been predicted to be uniquely stabilized by intra- and inter-plane super exchange and anti-symmetric DM interactions. Any changes to this delicate balance of interactions leads to the possibility of multiple magnetic phase's formation with a complex behavior. The various types of defects that can change the magnetic structure which will alter the exchange interactions are;

1. Anti-site defects such as exchange of cations belonging to different sites;
2. Truncation of the helical spin order leading to the formation of weak ferromagnetic entities; and
3. Non-magnetic grain boundary phases.

The anti-site defects can change the nature of super exchange interactions by altering the length and angle of the exchange path from antiferromagnetic to ferromagnetic as per the Goodenough-Kanamori-Anderson rule.[39-41] In the case of BNFSO $Fe^{3+}$ should occupy the tertrahedral sites in the ordered crystal. However, it is known that $Fe^{3+}$ can occupy octahedral sites also and in such cases it will lead to formation of anti-site defects – $Fe^{3+}$ replacing $Nb^{5+}$. This leads to changing of the super exchange path length as well as the angle resulting in variation of the antiferromagnetic and ferromagnetic interactions. Recent studies have indeed shown that $Fe^{3+}$ can occupy two non-equivalent sites and also the structural transition can lead to the formation of either a monoclinic symmetry or just result in reducing the symmetry to stay in the trigonal class with P3 symmetry.[26,42] Interestingly, substitution of $Ba^{2+}$ with $Sr^{2+}$ changes the magneto-electric behavior completely. Only one phase, magnetic and electric, is present at any given T and H. Substitution leads to the formation of a ordered structure with no intermediate magnetic transitions. Electrically however the system still shows a relaxor like behavior before it transforms into a long range ordered ferroelectric phase at $T_N$.

The above described magnetic and electric behaviors of BNFSO and SNFSO can be summarized by a phenomenological phase stability diagram as shown in Figure 10. Magnetically BNFSO exists in multiple electronically phase separated states at any temperature in the range 5 K to 300 K. For T > 100 K, paramagnetic, ferromagnetic and antiferromagnetic phases exist in the system. The isothermal magnetic hysteresis combined with ac magnetic susceptibility studies point to the existence of these phases. On cooling to 40 K < T < 100 K, the composition of the magnetic phases changes. A cluster glass phase is present together with ferromagnetic and antiferromagnetic phases in this temperature range. On further cooling to T < 40 K the system transforms to a mixture of antiferromagnetic and ferromagnetic phases. The antiferromagnetic phase present at all temperatures provides the exchange bias to the ferromagnetic phase. The presence of multiple phases is plausibly due to entities such as anti-site defects and termination of the helical magnetic structure into a partial helix as shown on the top in Figure 10. In comparison to the complex magnetic phases distribution in BNFSO, the scenario in the case of SNFSO is simple and 'clean'. The system exists either in the paramagnetic state above $T_N$ or antiferromagnetic state below $T_N$. A complete absence of multiple phases clearly shows that a smaller $Sr^{2+}$ ion in the same dodecahedral site results in significant magnetic structural stabilization. The dielectric studies on the other hand for the two compounds are relatively similar. In the case of BNFSO a true ferroelectric phase transition is not clearly observed. The system transforms to a low temperature relaxor phase from the high temperature paraelectric phase via a second relaxor phase. These results are in complete agreement with the magnetic results which exhibit complex phenomena. The Sr-substituted compound on the other hand exhibits a relatively simple transition into the ferroelectric state at low temperatures mediated by the relaxor phase. In both the compounds the dielectric studies reflect the magnetic results. The similarity in either the complexity observed in BNFSO or the simplicity observed in SNFSO clearly illustrate the strong coupling between magnetic and electric orders in the two compounds.

**Conclusions:**

The magnetic and dielectric properties of Fe-substituted langasite compounds $Ba_3NbFe_3Si_2O_{14}$ and $Sr_3NbFe_3Si_2O_{14}$ have been investigated in detail using a combination of structural and physical properties characterization down to 5 K. The main difference between the two compounds is the alkaline earth metal cation – $Sr^{2+}$ and $Ba^{2+}$ which are isovalent but with different size. The two compounds exhibit a completely different magnetic structure and behavior and hence the coupled dielectric behavior. The difference in behavior is plausibly due to the differences in super-exchange path length and angle which can stabilize either the antiferromagnetic or ferromagnetic interactions. Also, in the case of $Ba_3NbFe_3Si_2O_{14}$ the presence of antisite defects and incomplete magnetic helical arrangement have been found to result in the formation of multiple magnetic phases. These results clearly illustrate the role of frustration – geometric and magnetic, on the presence of multiple ground states which lead to complex phase separation phenomena. The proposed phenomenological magnetic and electric structures for the two compounds require a more exhaustive theoretical and experimental investigation using techniques such as temperature dependent magnetic and scanning microwave microscopies. Currently efforts are being made to perform these studies.

**Acknowledgements:**

The authors wish to acknowledge the Central facilities of Indian Institute of Technology Bombay for Magnetic Measurements and the Bhabha Atomic Research Centre for low temperature dielectric and neutron diffraction measurement facilities.

**Table 1:** The crystallographic parameters obtained from refinement of room temperature diffraction pattern for $Ba_3NbFe_3Si_2O_{14}$ ($R_{Bragg}$ = 2.0; $R_{WP}$ = 2.6; $\chi^2$ =1.4) and $Sr_3NbFe_3Si_2O_{14}$ ($R_{Bragg}$ = 15; $R_{WP}$ = 20; $\chi^2$ = 2.88) respectively. The exchange path for strongest magnetic interaction $J_1$ (*, in plane) and $J_5$ (+, out of plane) are also given at the bottom of the table. The structural parameters for $Sr_3NbFe_3Si_2O_{14}$ are shown in parentheses, [ ].

| Atom | Wyckoff | x | y | z | Occup. |
|---|---|---|---|---|---|
| Ba (Sr) | 3e | 0.4347(6) [0.4340(3)] | 0 | 0 | 0.5 |
| Nb | 1a | 0 | 0 | 0 | 0.166 |
| Fe | 3f | 0.7525(3) [0.7531(5)] | 0 | 0.5 | 0.5 |
| Si | 2d | 0.333 | 0.666 | 0.4828(1) [0.5170(4)] | 0.333 |
| O1 | 2d | 0.333 | 0.666 | 0.7769(2) [0.850(6)] | 0.333 |
| O2 | 6g | 0.4728(5) [0.498(2)] | 0.2984(5) [0.8238(6))] | 0.6471(8) [0.675(3)] | 1.00 |
| O3 | 6g | 0.2142(5) [0.2609(6)] | 0.0997(5) [0.0916(7)] | 0.2233(6) [0.739(3)] | 1.00 |

| *Fe-O3-O3-Fe, Å | + Fe-O3-O3-Fe, Å | a, Å | c, Å |
|---|---|---|---|
| 6.43 (2) | 6.447(5) | 8.5001(1) | 5.2267(1) |
| [6.42(3)] | [6.621(3)] | [8.2605(2)] | [5.1298(1)] |

**Figure Captions:**

**Figure 1:** Schematic diagram showing the arrangement of various cations in the Fe-substituted langasite compound $(Ba/Sr)_3NbFe_3Si_2O_{14}$. The ab-plane magnetic interactions $J_1$ and $J_2$ and the intra-plane interactions $J_3$, $J_4$ and $J_5$ which result in the formation of trimer and helical spin structures are shown. The incommensurate structure along the c-direction has a length of 7c.

**Figure 2:** The room temperature X-ray diffraction patterns for BNFSO and SNFSO together with the neutron diffraction pattern for BNFSO are shown in (a). The structural refinement was performed using Rietveld refinement and the refined structure diffraction data is also shown together with the experimental data points. The microstructure revealed after thermal etching at 900 °C clearly shows large grains as seen in the scanning electron micrograph (b). The average grain size is ~ 5μm with little variation in the size.

**Figure 3**: The core level XPS spectra for constituent elements in $Ba_3NbFe_3Si_2O_{14}$: Ba *3d* (**a**), Nb *3d* (**b**), Fe *2p* (**c**), Si *2p*(**d**), and O *1s*(**e**); the symbols represent the experimental data and solid line shows fitting. The satellite peaks are marked as S. The XPS spectra confirm that all the elements are in their expected oxidation states. (**f**) The electron spin resonance spectrum at room temperature also confirms that magnetic ion $Fe^{3+}$ to be in the spin only configuration with S = 5/2.

**Figure 4**: The powder neutron diffractograms obtained from $Ba_3NbFe_3Si_2O_4$ at different temperatures, below and above $T_N$. The structural refinement clearly shows the presence of a single phase at all T and the refined data together with the experimental data is shown in (a). The unique incommensurate magnetic structure peaks are marked with * and the peaks that are common to the magnetic and crystal structure are marked with + in the 6 K diffractogram. The variation of the two lattice parameters `a' and `c' with T, shown in (b), has a clear peak at the transition temperature.

**Figure 5:** The variation of specific heat capacity $C_p$ both as a function of temperature T and magnetic field H shows a clear peak at $T_N$ with a broad hump in the range 40 K to 100 K in (a) BNFSO and (b) SNFSO. The hump signals the onset of magnetic short range order with the peak indicating the formation of long range antiferromagnetic entities.

**Figure 6:** The variation of specific magnetization M with T and H shows path dependent irreversibility in BNFSO, (a). The field cooled (FC) and zero field cooled (ZFC)

magnetizations diverge for T < $T_{irr}$ and $T_{irr}$ decreases with increasing H, a behavior typical of magnetic glasses. The transition at 26 K, $T_N$, however is independent of `H'. In the case of SNFSO (b) however the magnetization does not exhibit any irreversibility with only a phase transition at 25 K, $T_N$ at all fields.

**Figure 7**: The real part of the ac magnetic susceptibility measured as a function of frequency ω and temperature T of BNFSO (a), shows a frequency independent peak at 26 K, $T_N$ and a frequency dependent dispersive peak in the range 40K to 90K. The temperature dependence of the peak frequency fitted to the critical slowing dynamics model, power law dependence, is shown in the inset. The susceptibility of SNFSO on the other hand shows a single, non-dispersive peak at 25 K with no other peaks, (b).

**Figure 8 :** The hysteresis of specific magnetization M with field `H' and a non-saturating high field behaviour at constant temperature in the case of BNFSO indicates presence of a ferromagnetic component at all temperatures along with a antiferromagnetic component (a). The specific magnetization of SNFSO however does not show any hysteresis indicating the presence of a single phase at any T and H, (b). The derivative of the specific magnetization dM/dH as a function of H clearly illustrates the hysteretic nature, (c) and (d). The coercivity $H_c$ also exhibits exchange bias, ΔH which changes sign below $T_N$ and above $T_N$ showing the presence of an antiferromagnetic component together with the ferromagnetic component in BNFSO is shown in the inset of (a).

**Figure 9 :** The imaginary component of the dielectric constant $\epsilon''$ measured as a function of frequency ω in the range $10^1$ to $10^5$ Hz from room temperature down to 5K is shown in (a) and (b) for BNFSO and SNFSO respectively. In the case of BNFSO, $\epsilon''$ clearly shows two dispersive peaks, peak-I and peak –II in the temperature ranges 20 K to 35 K and 40 K to 125 K respectively, (a). There is only one high temperature dispersive peak in SNFSO, (b). In both the cases however $\omega^e_{max}$ has a Vogel Fulcher temperature dependence as seen in the insets.

**Figure 10 :** A phenomenological representation of magnetic and electric structures present at different temperatures in BNFSO (a) and SNFSO (b). The incomplete magnetic helical structure and modification of the in-plane super-exchange interaction $J_1$ due to an antisite defect are shown schematically on the top. AF, FM, CG, PM and PE represent antiferromagnetic, ferromagnetic, cluster glass, paramagnetic and paraelectric phases.

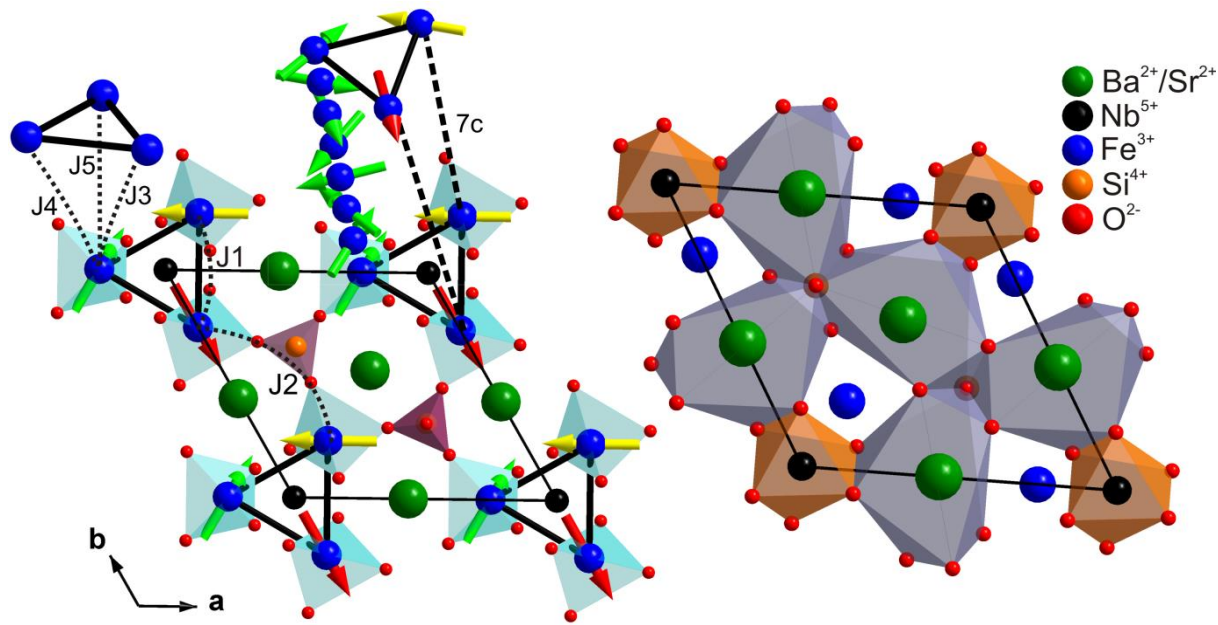

**Figure 1:** Schematic diagram showing the arrangement of various cations in the Fe-substituted langasite compound $(Ba/Sr)_3NbFe_3Si_2O_{14}$. The ab-plane magnetic interactions $J_1$ and $J_2$ and the intra-plane interactions $J_3$, $J_4$ and $J_5$ which result in the formation of trimer and helical spin structures are shown. The incommensurate structure along the c-direction has a length of 7c.

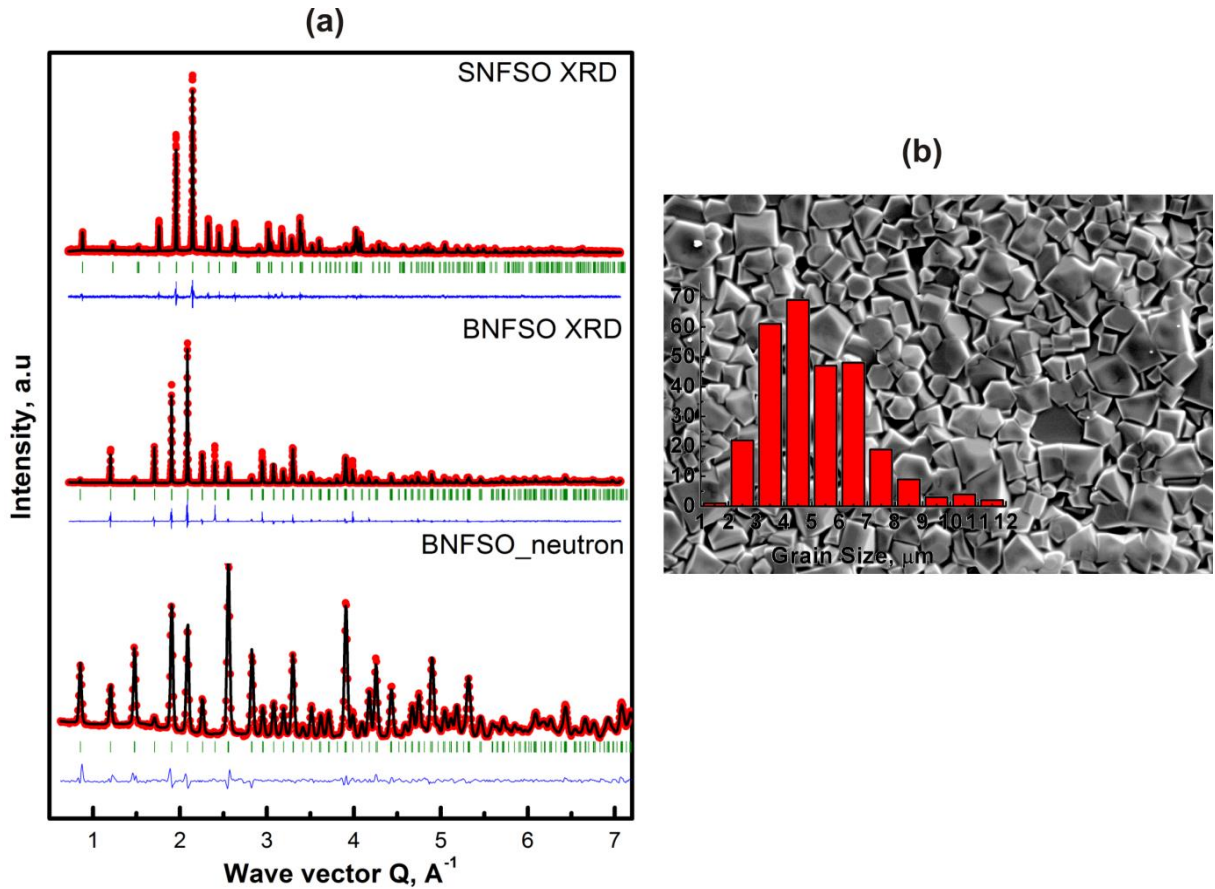

**Figure 2:** The room temperature X-ray diffraction patterns for BNFSO and SNFSO together with the neutron diffraction pattern for BNFSO are shown in (a). The structural refinement was performed using Rietveld refinement and the refined structure diffraction data is also shown together with the experimental data points. The microstructure revealed after thermal etching at 900 ºC clearly shows large grains as seen in the scanning electron micrograph (b). The average grain size is ~ 5 μm with little variation in the size.

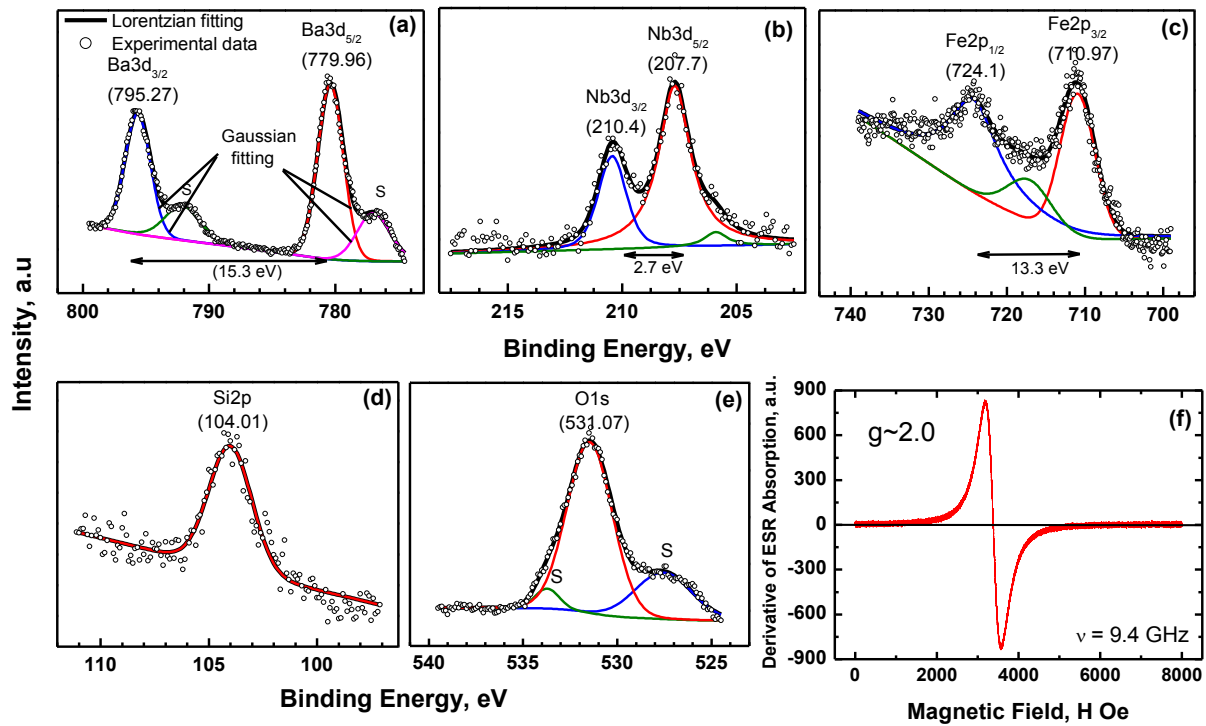

**Figure 3**: The core level XPS spectra for constituent elements in $Ba_3NbFe_3Si_2O_{14}$ : Ba *3d* (**a**), Nb *3d* (**b**), Fe *2p* (**c**), Si *2p*(**d**), and O *1s*(**e**); the symbols represent the experimental data and solid line shows fitting. The satellite peaks are marked as S. The XPS spectra confirm that all the elements are in their expected oxidation states. (**f**) The electron spin resonance spectrum at room temperature also confirms that magnetic ion $Fe^{3+}$ to be in the spin only configuration with S = 5/2.

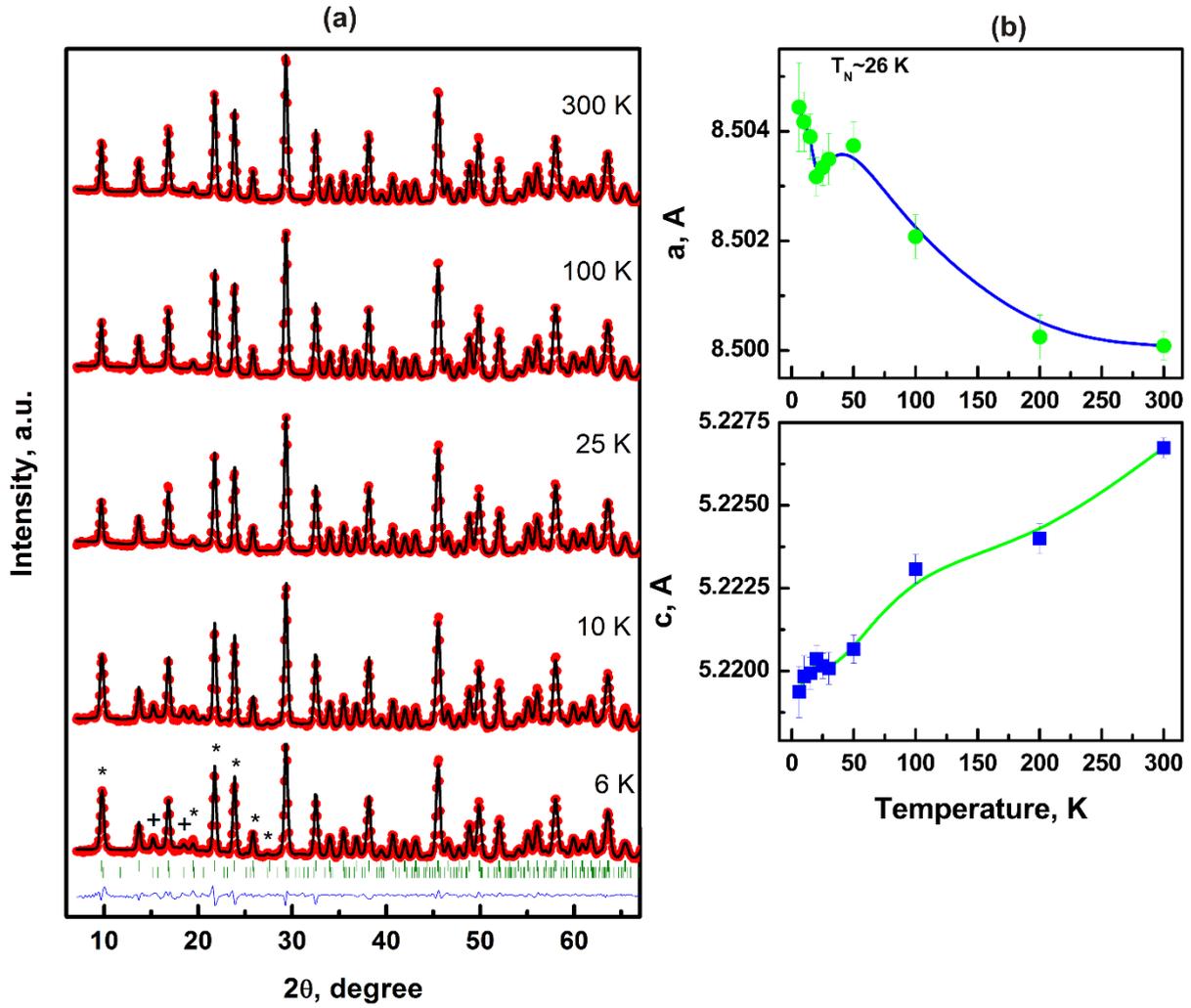

**Figure 4**: The powder neutron diffractograms obtained from $Ba_3NbFe_3Si_2O_4$ at different temperatures, below and above $T_N$. The structural refinement clearly shows the presence of a single phase at all T and the refined data together with the experimental data is shown in (a). The unique incommensurate magnetic structure peaks are marked with * and the peaks that are common to the magnetic and crystal structure are marked with + in the 6 K diffractogram. The variation of the two lattice parameters `a' and `c' with T, shown in (b), has a clear peak at the transition temperature.

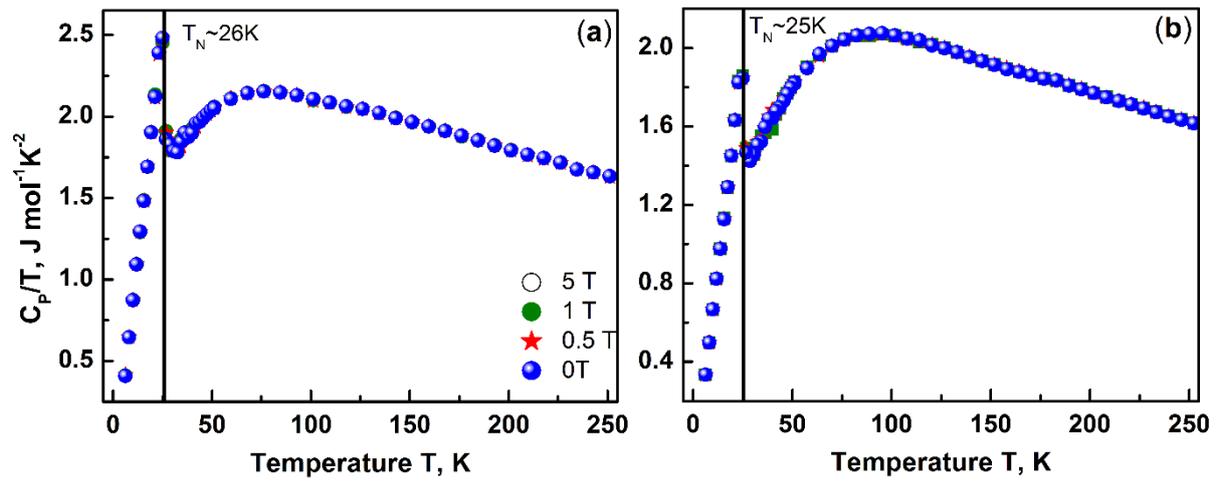

**Figure 5:** The variation of specific heat capacity $C_p$ both as a function of temperature T and magnetic field H shows a clear peak at $T_N$ with a broad hump in the range 40 K to 100 K in (a) BNFSO and (b) SNFSO. The hump signals the onset of magnetic short range order with the peak indicating the formation of long range antiferromagnetic entities.

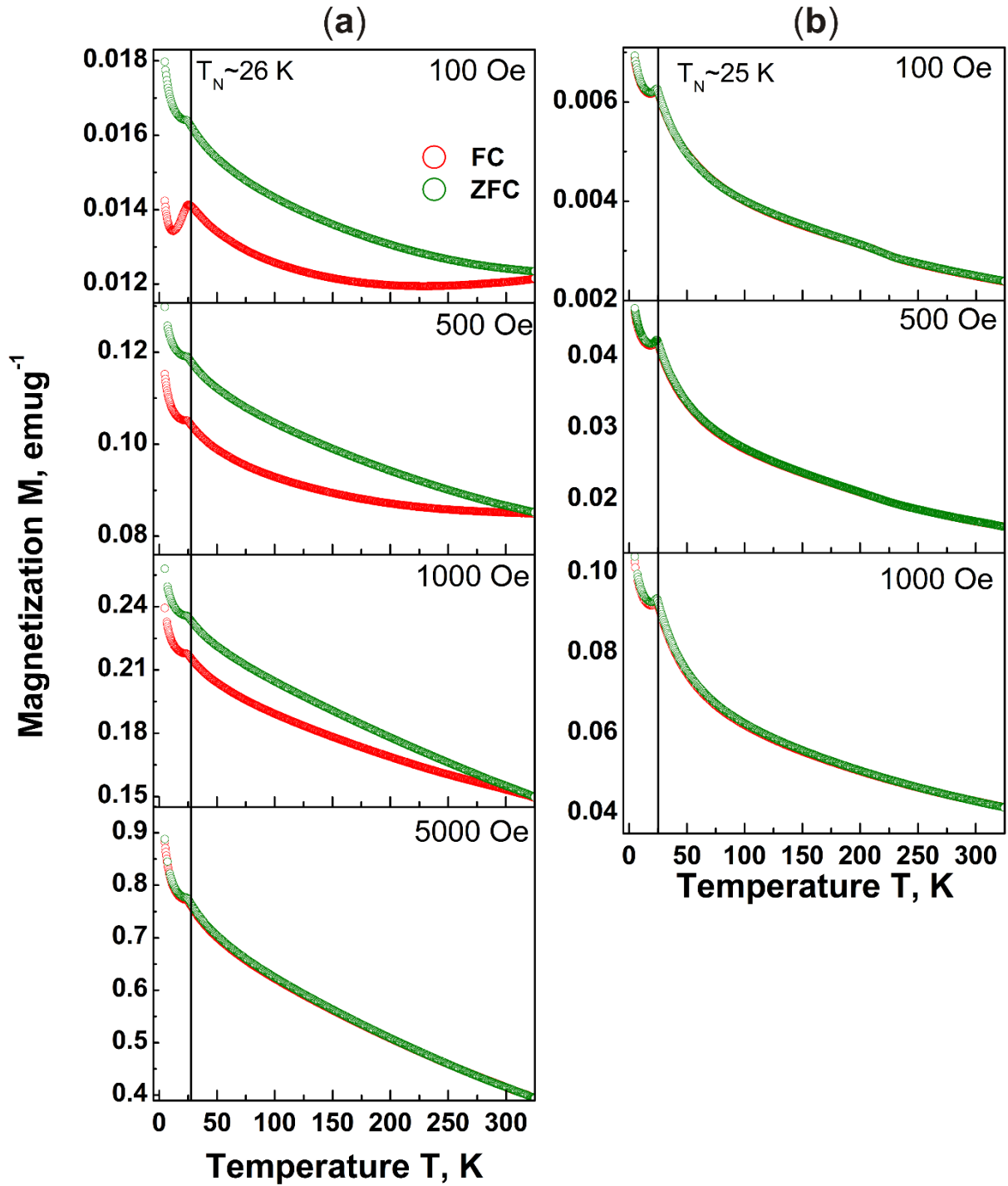

**Figure 6:** The variation of specific magnetization M with T and H shows path dependent irreversibility in BNFSO, (a). The field cooled (FC) and zero field cooled (ZFC) magnetizations diverge for $T < T_{irr}$ and $T_{irr}$ decreases with increasing H, a behavior typical of magnetic glasses. The transition at 26 K, $T_N$, however is independent of `H'. In the case of SNFSO (b) the magnetization does not exhibit any irreversibility with only a phase transition at 25 K, $T_N$ at all fields.

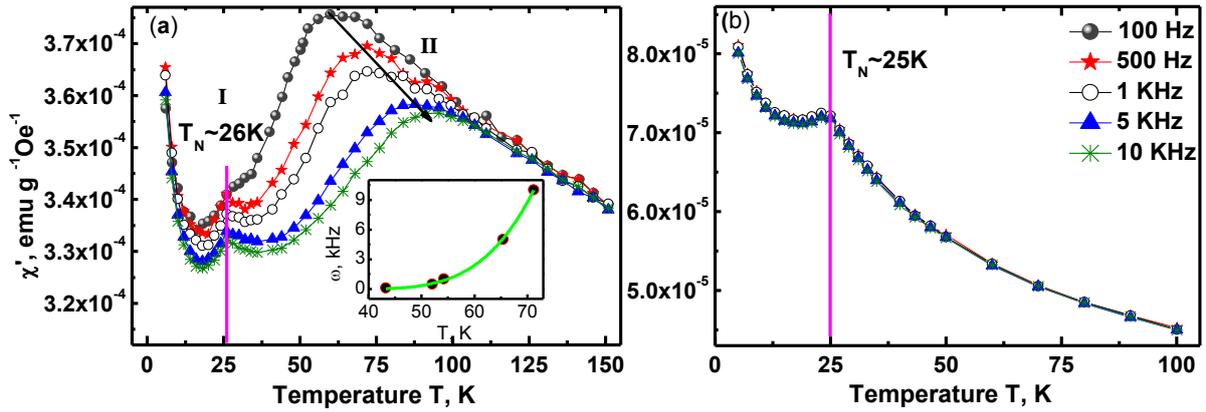

**Figure 7**: The real part of the ac magnetic susceptibility measured as a function of frequency ω and temperature T of BNFSO (a), shows a frequency independent peak at 26 K, $T_N$ and a frequency dependent dispersive peak in the range 40K to 90K. The temperature dependence of the peak frequency fitted to the critical slowing dynamics model, power law dependence, is shown in the inset. The susceptibility of SNFSO on the other hand shows a single, non-dispersive peak at 25 K with no other peaks, (b).

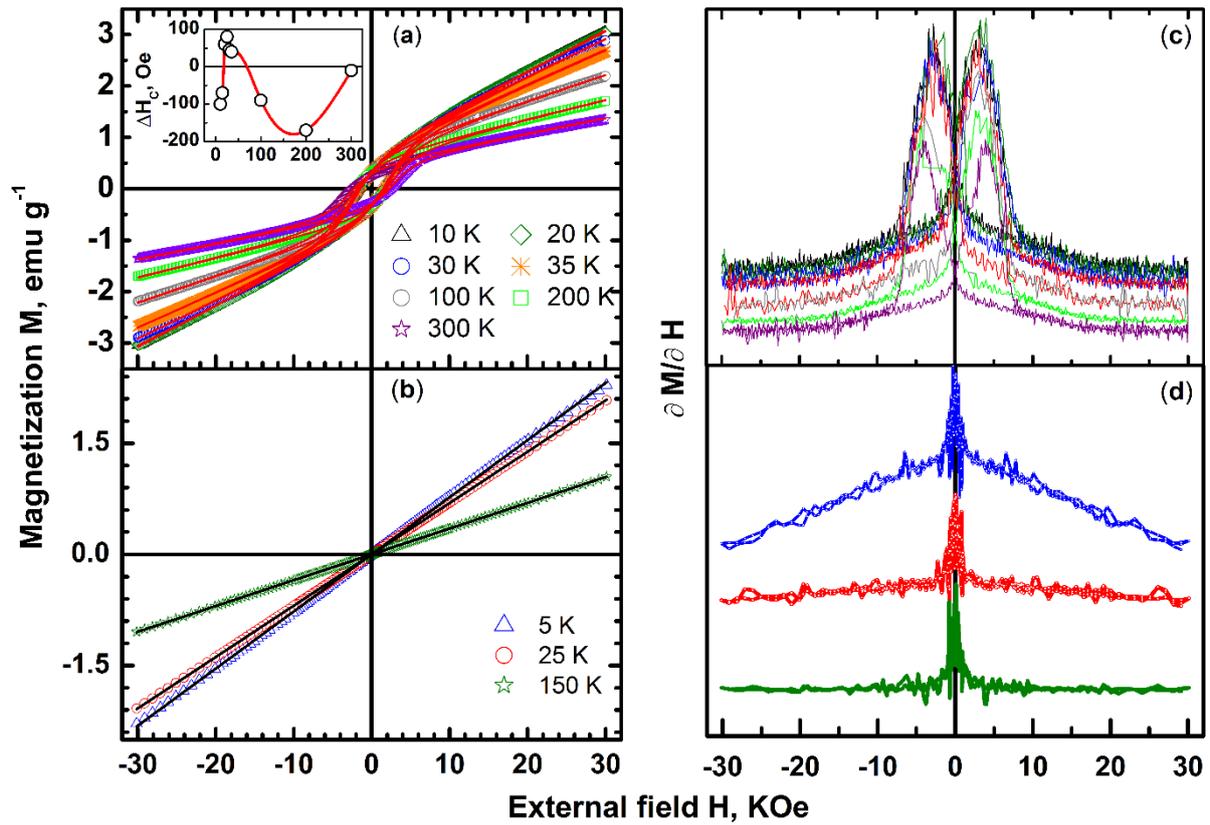

**Figure 8:** The hysteresis of specific magnetization M with field H and a non-saturating high field behaviour at constant temperature in the case of BNFSO indicates presence of a ferromagnetic component at all temperatures along with a antiferromagnetic component (a). The specific magnetization of SNFSO however does not show any hysteresis indicating the presence of a single phase at any T and H, (b). The derivative of the specific magnetization dM/dH as a function of H clearly illustrates the hysteretic nature, (c) and (d). The coercivity $H_c$ also exhibits exchange bias, $\Delta H$ which changes sign below $T_N$ and above $T_N$ showing the presence of an antiferromagnetic component together with the ferromagnetic component in BNFSO and is shown in the inset of (a).

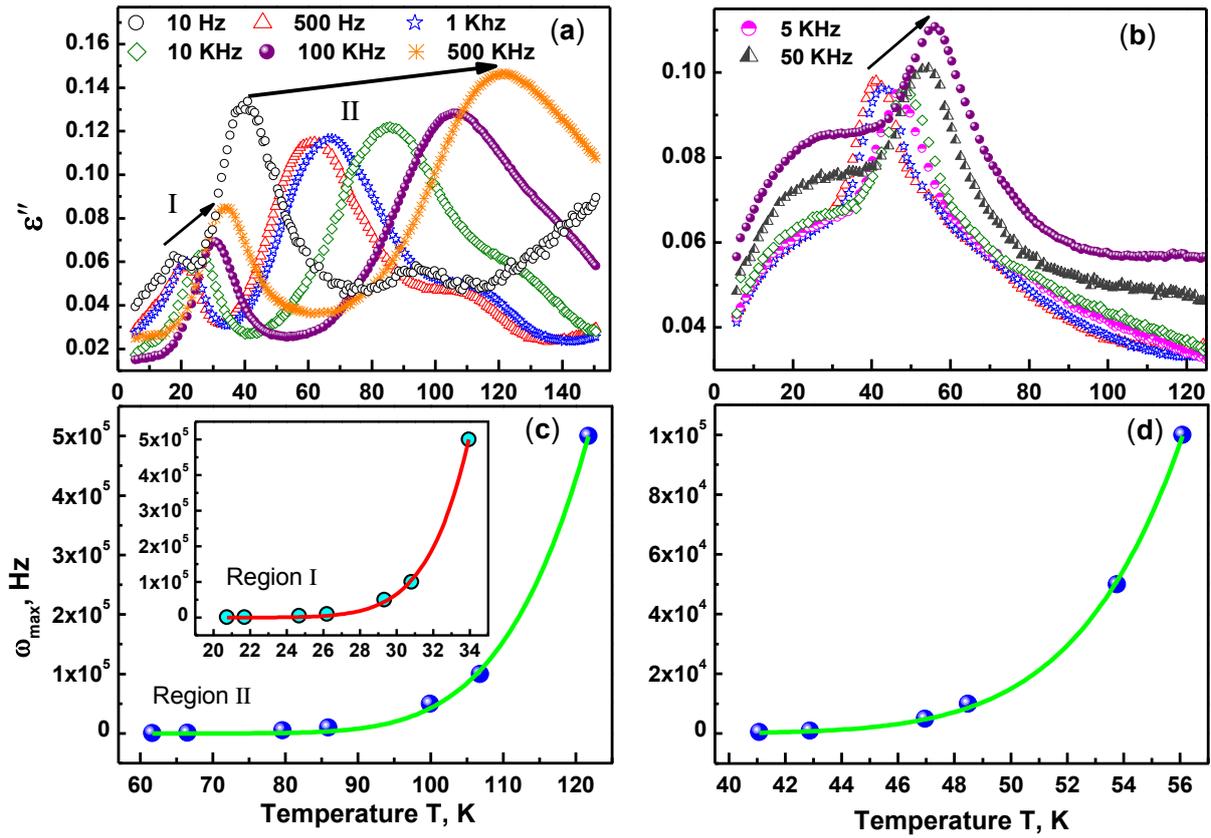

**Figure 9:** The imaginary component of the dielectric constant $\epsilon''$ measured as a function of frequency $\omega$ in the range $10^1$ to $10^5$ Hz from room temperature down to 5K is shown in (a) and (b) for BNFSO and SNFSO respectively. In the case of BNFSO, $\epsilon''$ clearly shows two dispersive peaks, peak-I and peak –II in the temperature ranges 20 K to 35 K and 40 K to 125 K respectively, (a). There is only one high temperature dispersive peak in SNFSO, (b). In both the cases however $\omega^e_{max}$ has a Vogel Fulcher temperature dependence as seen in (c) and (d).

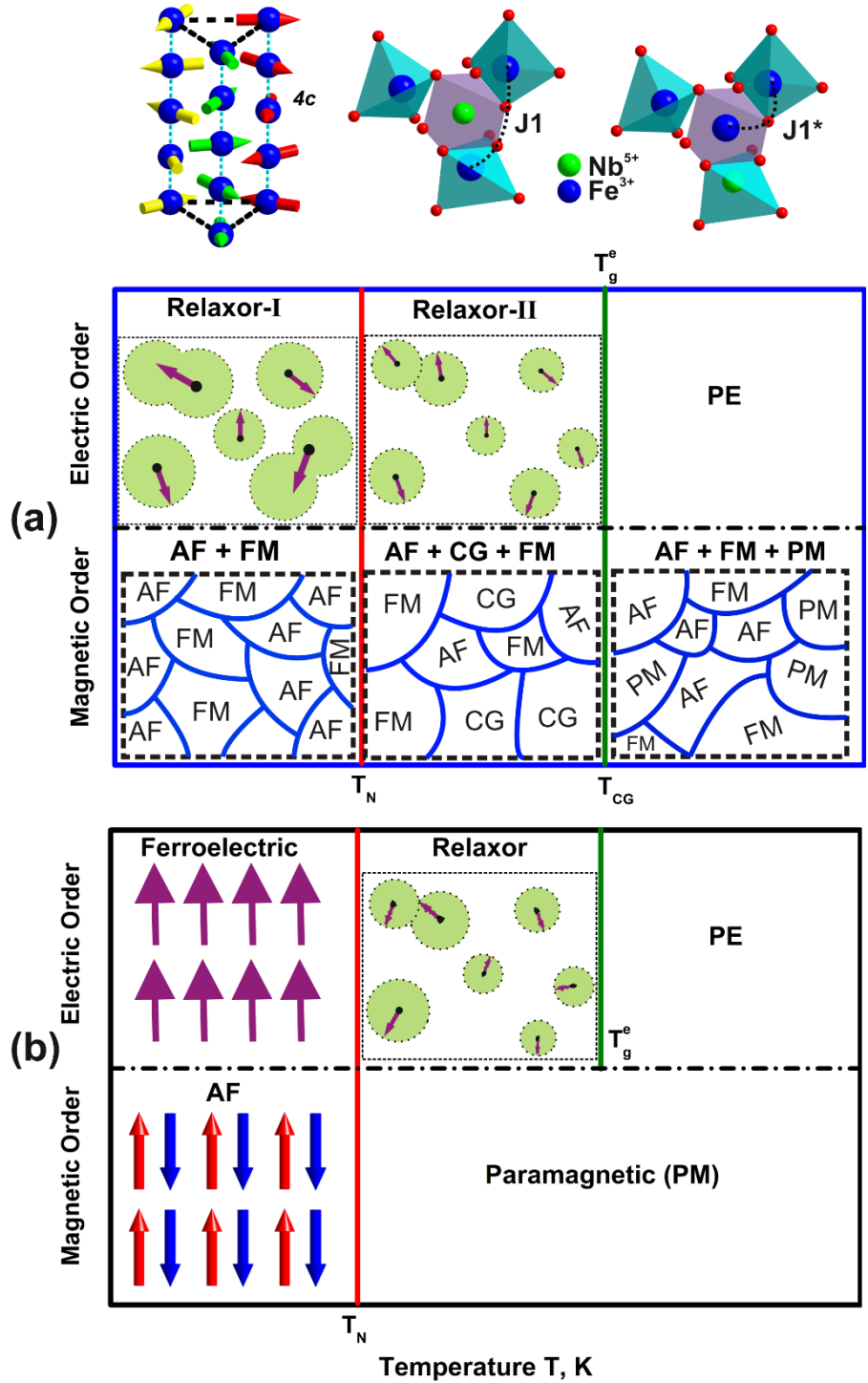

**Figure 10:** A phenomenological representation of magnetic and electric structures present at different temperatures in BNFSO (a) and SNFSO (b). The incomplete magnetic helical structure and modification of the in-plane super-exchange interaction $J_1$ due to an antisite defect are shown schematically on the top. AF, FM, CG, PM and PE represent antiferromagnetic, ferromagnetic, cluster glass, paramagnetic and paraelectric phases.

28